\documentclass[aps,letterpaper,11pt,nofootinbib,preprint,showpacs,preprintnumbers,amsmath,amssymb]{revtex4-1}%
\usepackage{latexsym}%
\usepackage{epsf}%
\usepackage[hypertex]{hyperref}%
\usepackage{graphicx}
\usepackage{epsfig}%
\usepackage{graphicx}%
\usepackage[lofdepth,lotdepth]{subfig}%
\usepackage{color}

\def\tsigma{{\widetilde\sigma}}

\newcommand{\beq}{\begin{equation}}
\newcommand{\beqn}{\begin{equation}\nonumber}
\newcommand{\eeq}{\end{equation}}
\newcommand{\bea}{\begin{eqnarray}}
\newcommand{\bean}{\begin{eqnarray}\nonumber}
\newcommand{\eea}{\end{eqnarray}}

\begin{document}
\begin{center}
{\bf{\Large Statistical analysis of entropy correction from topological defects in Loop Black Holes}}
\bigskip\bigskip

{{Kinjalk Lochan$^{a,}$\footnote{e-mail address: \tt{kinjalk@tifr.res.in}} and Cenalo Vaz$^{b,}$\footnote{e-mail address: \tt{cenalo.vaz@uc.edu}},
}}
\bigskip\bigskip

{\it$^a$Tata Institute of Fundamental Research,\\ Homi Bhabha Road, Mumbai-400005, India\\  
\bigskip
$^b$Department of Physics,}
{\it University of Cincinnati,}\\
{\it Cincinnati, Ohio 45221-0011, USA}
\end{center}
\bigskip\bigskip\bigskip\bigskip
\medskip

\centerline{ABSTRACT}
\bigskip\bigskip

In this paper we discuss the entropy of quantum black holes in the LQG formalism when the number 
of punctures on the horizon is treated as a quantum hair, that is we compute the black hole entropy 
in the grand canonical (area) ensemble. The entropy is a function of both the average area and
the average number of punctures and bears little resemblance to the Bekenstein-Hawking entropy. In 
the thermodynamic limit, both the ``temperature'' and the chemical potential can be shown to be functions 
only of the average area per puncture. At a fixed temperature, the average number of punctures 
becomes proportional to the average area and we recover the Bekenstein-Hawking area-entropy law to 
leading order provided that the Barbero-Immirzi parameter, $\gamma$, is appropriately fixed. This 
also relates the chemical potential to $\gamma$. We obtain a sub-leading correction, which differs 
in signature from that obtained in the microcanonical and canonical ensembles in its sign but agrees 
with earlier results in the grand canonical ensemble. 

\vfill\eject

\section{Introduction}
Loop quantum gravity (LQG) is a background independent non-perturbative canonical quantization 
of general relativity (GR) \cite{LQG} employing a canonical chart consisting of the $su(2)$ 
Lie algebra valued  Ashtekar-Barbero connection \cite{Connection} and densitized triads as canonical 
variables. Because of the geometric interpretation of these variables it is possible to construct 
operators in the quantum theory corresponding to classical geometric quantities such as 
area and volume. At the kinematic level, these operators have discrete spectra, quantized in units of 
an appropriate power of the Planck length (upto an arbitrary constant, the Barbero-Immirizi 
parameter). At the dynamical level, the question of finding the quantum Einstein's equations becomes
the problem of finding the state space lying in the kernel of the Hamiltonian constraint. Although a
solution to this problem remains to be found and the full picture of quantum gravity is yet to
emerge, the structure of space-time revealed at the kinematical level provides an interesting 
approach to black-hole physics when the black hole horizon is modeled as an isolated horizon 
\cite{isolated horizon}. The isolated horizon is taken as an inner boundary of space-time, which 
gets threaded by links of eigenstates of geometric operators, namely the ``spin-networks''. Each 
puncture contributes some quantum of area according to the $su(2)$ label it carries, which provides a 
microscopic description of the horizon and its degrees of freedom. Standard statistical mechanics 
techniques may be used to determine the entropy associated with the horizon and the results can then be
tested against the semi-classical theory, in particular to the Bekenstein-Hawking (BH) entropy 
\cite{B-H law}. This program has been carried out successfully in LQG. The BH entropy is recovered as 
a dominant term at the macroscopic level and sub-dominant terms of quantum origin are also 
obtained \cite{Counting entropy, gm06}.

It is worth noting that the BH area-entropy law has also been obtained in diverse approaches to quantum 
gravity and each of these approaches has succeeded in providing an alternative interpretation of the quantum 
gravitational degrees of freedom comprising a black hole. Sub-dominant, logarithmic corrections to the 
BH law have also been proposed in these approaches \cite{logterms}. 

In LQG the black hole is traditionally treated within the microcanonical ensemble, since the horizon 
is taken to be an isolated horizon \cite{IH}. Use of the microcanonical ensemble is generally justified 
by the fact that because there is no energy flux across the isolated horizon, the area is expected 
to remain constant. However, when the possibility of black hole evaporation is taken into account,
an exchange of the number of punctures or the spin-label of a puncture with the bulk geometry can, 
in principle, give rise to area fluctuations as well as fluctuations in the number of punctures. 
Thus it appears that the canonical and/or grand canonical ensembles, in which the number of punctures 
is treated as a statistical variable, are more appropriate choices if one intends eventually 
to describe physical processes involving black hole evaporation. There is another reason why 
the microcanonical ensemble may not be suitable for the study of black hole thermodynamics: it is 
now well known that the microcanonical entropy, apart from being non-uniquely defined, is not a differentiable 
function of the area but rather a staircase (or ladder) function \cite{cdf07,pp11}. As pointed out in 
\cite{bv11a}, this makes it difficult to interpret basic thermodynamic variables, such as the temperature, 
which are defined in terms of derivatives of the entropy. The canonical and grand canonical partition 
functions are assumed, by contrast, to be smooth functions of their arguments. Therefore, recently there 
has been renewed interest in treating the horizon within the canonical and grand canonical area ensembles 
\cite{Ghosh,Barbero}. 

The idea of using the horizon area as the ensemble variable, as opposed to the more traditional energy 
ensembles, dates back to the work of Krasnov \cite{Krasnov}. It was originally motivated by the fact 
that if a microscopic description of the black hole entropy is to yield something like the BH law then 
the density of states should satisfy a relation of the form $\ln\Omega(M) \sim M^2$, which 
would lead to a divergent partition function in the energy ensemble. A novel justification for 
its use may be found in \cite{Frodden}. The use of the number of punctures, $N$, as a statistical 
variable was advocated by \cite{Major} and recently revived in \cite{Ghosh}.

In \cite{ours} we presented a detailed analysis of black holes with the LQG area spectrum in the 
canonical area ensemble (fixed number of punctures) and obtained a logarithmic correction with a
negative coefficient, in agreement with the microcanonical entropy results. In \cite{bv11a}, 
the entropy was computed in the grand canonical ensemble for two counting schemes, {\it viz.,} the one 
proposed by Domagala and Lewandowski (DL) \cite{dl04} and another, more recently proposed by Engle, 
Noui and Perez (ENP) \cite{enp10}. The chemical potential was set to zero from the start and a 
sub-dominant logarithmic correction with a positive coefficient was determined in the DL case whereas 
no correction was obtained in the ENP case. The sign of the correction is important because a positive 
sign (as in the case of DL) makes the entropy a concave function of the area whereas it is a convex 
function if the sign is negative or if there is no correction (as in the case of ENP). Here we also 
work with the grand canonical ensemble, but we do not set the chemical potential to zero.  We employ 
the counting scheme originally proposed by Ghosh and Mitra (GM) \cite{gm06}, but our methods are 
applicable to the DL scheme as well. Because the chemical potential is not set to zero from the start, 
the entropy ends up being a function of the mean area, $A$, and the mean number of punctures, $N$,  
making it clear that the positive sign attached to the logarithmic correction is due to a contribution 
coming from fluctuations in $N$.   In the large $N$ limit, both the ``inverse temperature'' (by which 
we will mean the variable conjugate to the horizon area) and the chemical potential (variable conjugate 
to $N$) are functions {\it only} of the mean area per puncture and so they are related to one another 
as well. At a constant temperature $A$ becomes proportional to $N$ and only in this case does one 
obtain an entropy-area relation whose dominant contribution agrees with the BH law and whose sub-dominant 
contribution is identical to the results of \cite{bv11a} for the DL counting scheme. 

We first look at the grand canonical emsemble for isolated horizons, {\it i.e.,} without the 
projection constraint, in section II. This will set the stage for later calculations and here we 
will show that fluctuations in $N$ lead to an additional logarithmic correction to the entropy, 
coming with a coefficient $+1$. In Section III we introduce the projection constraint and evaluate 
the partition function by the saddle point method in the large $N$ (thermodynamic) limit. We 
determine the entropy of a black hole horizon as a function of $A$ and $N$ and obtain a logarithmic 
correction to the BH entropy at constant temperature, but it comes with a positive coefficient of 
$+1/2$, in agreement with the results of \cite{bv11a} for the DL counting scheme. Thus the effect of the 
projection constraint is to reduce the number of configurations by subtracting $\frac 12\ln N$ 
from the expression for the entropy of an isolated horizon. This was also determined in \cite{ours} 
for the canonical ensemble. In section IV we verify the validity of the saddle point method by comparing 
its results with the exact partition function for an equispaced area spectrum. We conclude in Section 
V with a brief summary of our results and a short discussion of the outlook for future work.

\section{The Grand Canonical Ensemble}

We start with a generic isolated horizon and refrain from applying the projection constraint. We 
want to construct the grand canonical partition function for this system in the GM counting scheme, 
so let us begin with 
\beq
\Xi(\beta,\alpha) = \sum_{N=0}^\infty\sum_{n_j=0}^N \frac{N!}{\prod_j n_j!} \prod_j (2j+1)^{n_j} 
e^{-(8\pi \gamma\beta a_j -\alpha)n_j},
\eeq
(as in \cite{ours}) where the first term represents the degeneracy of states with $N$ punctures, $n_j$ 
of which contribute area $a_j=l_p^2\sqrt{j(j+1)}$ with $j \in\{1/2,1,3/2,\ldots\}$ and $(2j+1)$ is the 
degeneracy associated with spin $j$. The Barbero-Immirzi parameter is $\gamma$ and the variables $\beta$ and 
$\alpha$ are Lagrange multipliers, which can be thought of as an ``inverse temperature'' conjugate to the area 
and a ``chemical potential'' conjugate to the number of punctures respectively (it is $\mu=\alpha/\beta$ 
that is conventionally called the chemical potential, but we shall simply use $\alpha$). 

Summing over the punctures,
\beq
\Xi(\beta,\alpha) = \sum_{N=0}^\infty \left(\sum_j (2j+1) e^{-(8\pi \gamma\beta a_j -\alpha)}
\right)^N = \frac 1{1-\lambda(\alpha) z(\beta)},
\label{Pf}
\eeq
where $\lambda(\alpha) = e^\alpha$ is the fugacity, and
\beq
z(\beta) = \sum_j (2j+1) e^{-8\pi \gamma\beta a_j}.
\eeq
The average occupation number of punctures  in a state $j$ will be 
\beq
\langle n_j \rangle = -\frac 1{8\pi\gamma\beta} \frac{\partial \ln\Xi}{\partial a_j} = 
\frac{\lambda (2j+1) e^{-8\pi\gamma\beta a_j}}{1-\lambda z},
\eeq
and the average total number of punctures is given by
\beq
\langle N \rangle = \frac{\partial \ln \Xi}{\partial \alpha} = \sum_j \langle n_j\rangle = 
\frac{\lambda z}{1-\lambda z}.
\label{nump}
\eeq
Non-negativity of $\langle N \rangle$ means that $\lambda z \lesssim 1$ and, furthermore, the average 
area of the horizon will be given by (henceforth dropping the angular brackets)
\beq
A = - \frac{\partial\ln\Xi}{\partial \beta} = \frac{\partial}{\partial\beta}
\ln (1-\lambda z) = -\frac{\lambda}{1-\lambda z} \frac{\partial z}{\partial\beta} = - \frac Nz 
\frac{\partial z}{\partial \beta} = - N \frac{\partial \ln z}{\partial \beta},
\eeq
or
\beq
\frac{A}{4\pi\gamma l_p^2 N} = - \frac{\partial \ln z(\sigma)}{\partial \sigma}~ \stackrel{\text{def}}
{=}~ a.
\label{qdef}
\eeq
where we set $\sigma=4\pi\gamma \beta l_p^2$. Solving this equation would give $\sigma=\sigma(a)$, 
{\it i.e.,} the ``temperature'' depends only on the average area per puncture, $a$. We recognize this 
as precisely equation (9) in \cite{ours}, therefore the relationship between $\sigma$ and $a$
remains intact as we go from the canonical to the grand ensemble and from \eqref{nump} we find the 
fugacity,
\beq
\lambda=\lambda(a,N) = \frac{N}{(N+1)z(a)}.
\label{lam}
\eeq
On the other hand from (\ref{Pf}) and \eqref{lam}, 
\beqn
\ln \Xi = \ln (N+1),
\eeq
so the Legendre transform of $\ln \Xi$, which is the entropy, becomes
\beq
S(A,N) = \ln \Xi + \beta A -\alpha N = (N+1)\ln (N+1) -N\ln N + Na\sigma(a)+  N\ln z(a)
\eeq
and simplifies, in the limit of large $N$, to 
\bea
S(A,N) &\approx& \ln N + N [a \sigma(a)+ \ln z(a)]\cr\cr
&=& \frac{\sigma(a)}{\pi\gamma} \frac A{4l_p^2} + N \ln z(a) + \ln N.
\label{san}
\eea
This can be compared with equation (12) of \cite{ours}. It is the general expression for the 
entropy of an isolated horizon of average area $A$ and carrying $N$ punctures.

First, we notice that allowing fluctuations in the number of punctures has introduced an additional 
logarithmic term in $N$. The fact that this term is missing in  the canonical ensemble \cite{ours} 
is an artifact of holding $N$ fixed. Secondly, the entropy in \eqref{san} is not the BH entropy in
general but a complicated function of $A$ and $N$. However, an important consequence of \eqref{qdef} 
and \eqref{lam} is that both $\sigma$ and  $\lambda$ depend exclusively on $a$ in the large $N$ limit 
therefore, for example, $a$ and $\lambda$ remain fixed during any isothermal process involving the 
isolated horizon. Suppose that at some fixed value of the temperature, $\sigma_0$, or of the chemical 
potential, $\alpha_0$, we find that $a(\sigma_0)=a_0$ then 
\beq
N = \frac A{4\pi\gamma l_p^2 a_0}
\eeq
can be used to eliminate $N$ in \eqref{san} and gives a reduced entropy function
\beq
S(A) \approx \frac 1{\pi\gamma}\left[\sigma_0+\frac{\ln z(a_0)}{a_0}\right]\frac A{4l_p^2} + \ln 
\frac A{4l_p^2} + \text{const.},
\label{simpleent}
\eeq
which is consistent with the BH law, but only so long as the temperature is held fixed. In this 
isothermal condition, agreement with the BH entropy to leading order then determines the 
Barbero-Immirzi parameter in terms of the temperature as well, according to 
\beq
\gamma = \frac 1{\pi}\left[\sigma_0+\frac{\ln z(a_0)}{a_0}\right].
\eeq
An additional criterion is required in order to fully determine all the parameters. For example, a 
possible condition for determining $\sigma_0$ may be taken to be a vanishing chemical potential, 
as in \cite{bv11a}, by which the punctures are effectively treated as a photon gas. The condition 
for zero chemical potential, according to \eqref{lam}, is
\beq
z(\sigma_0) \approx 1,
\label{theeq}
\eeq
as obtained in \cite{ours}, and it determines $a_0$. Solving \eqref{theeq} numerically one finds, as in
\cite{gm06}, that $\sigma_0 = 0.861$, $a_0=2.921$ and $\gamma=0.274$. 

If the chemical potential is zero it is important to understand how the punctures get ``thermalized''. 
In the analogous situation of a photon gas, the photons are thermalized by interactions with the walls 
of the container. It is possible that such a thermalization occurs via dynamical processes near the 
horizon, but it is also possible that such processes cause $\alpha$ to be different from zero. 

Note that in the above arguments it has not been necessary to obtain an explicit solution for $z(\sigma)$.
However, it is possible to obtain a closed form expression for $z(\sigma)$ in a certain approximation that 
will be useful in a subsequent section, when the projection constraint is implemented. In this approximation 
\cite{Rovelli} we replace $2j$ by $l$ so $l \in \{1,2,,\ldots\}$ and $a_j \rightarrow a_l=\frac 12 l_p^2\sqrt{l(l+2)} 
\approx \frac 12 l_p^2 (l+1)$. This is an excellent approximation for large values of $l$ but not so good 
for low values of $l$, so we introduce errors for low spin punctures. Although the mathematical steps to 
estimate the error in this approximation are discussed in \cite{ours}, it has been shown that a 
coherent state representation of the Schwarzschild horizon \cite{KSD} or a unitary invariant representation of 
the area operator \cite{Livine} lead to precisely this kind of spectrum.

For this equispaced spectrum we obtain
\beq
z^{(0)}(\sigma) = \frac{2-e^{-\sigma}}{(e^\sigma-1)^2}
\label{zapprox}
\eeq
for the partition function. The zero chemical potential condition in \eqref{theeq} then gives 
$\sigma_0 = 0.810$, $a_0=3.318$ and $\gamma=0.258$, values that are not far from the exact 
ones quoted above.

\section{Projection Constraint}

With the projection constraint, the {\it canonical} partition function for a fixed number of punctures 
was given in \cite{ours} as
\beq
Z(\sigma,N) = \int_{-\pi}^\pi \frac{dk}{2\pi} \left(\sum_{l=1}^\infty e^{-\sigma
\sqrt{l(l+2)}}~ \frac{\sin k(l+1)}{\sin k}\right)^N
\eeq
To obtain the grand canonical partition function, it is necessary to introduce a chemical potential 
and sum over $N$. Therefore, the projection constraint will lead to a partition function of the form
\beq
\Xi(\sigma,\alpha) =  \sum_N \int_{-\pi}^{\pi} \frac{dk}{2\pi} \left(\sum_{l=1}^\infty 
e^{-\sigma\sqrt{l(l+2)}}~ \frac{\sin k(l+1)}{\sin k}\right)^N e^{\alpha N}.
\eeq
and performing the sum over $N$ we arrive at
\beq
\Xi(\sigma,\alpha) = \frac 1{2\pi}\int_{-\pi}^\pi \frac{dk}{1-\lambda(\alpha) \sum_{l=1}^\infty z_l(\sigma) 
\left(\frac{\sin k(l+1)}{\sin k}\right)},
\label{partfunc}
\eeq
where we have set $z_l(\sigma) = e^{-\sigma\sqrt{l(l+2)}}$ and $\lambda(\alpha)=e^\alpha$.

It does not seem possible to perform the integral exactly, but it may be approximated by steepest 
descent. First we note that that the integrand is an unimodal, symmetric distribution centered at 
$k=0$ so long as the denominator of \eqref{partfunc} is positive. If $\lambda = 1$ then this is the 
condition that $\sigma > \sigma_0 = 0.861...$, which solves equation \eqref{theeq}. We then extend 
the integration to the entire real line by the change of variables 
\beqn
k(x) = 2 \tan^{-1}(x/2),
\eeq
under which the integrand becomes
\beq
f(x) = \frac 1{2\pi} \frac {dk(x)/dx}{1-\lambda(\alpha) \sum_{l=1}^\infty z_l(\sigma) \left(\frac{\sin k(x)
(l+1)} {\sin k(x)}\right)}
\eeq
and finally we assume that the distribution is sufficiently well approximated by a normal distribution 
(centred at $x=0$) in the limit of large $N$ and to leading order in $N$, so we replace $f(x)$ by
\beq
f(x) \approx  f(0) \exp\left[-\frac{x^2}{2\tsigma^2}\right].
\eeq
The variance is given by
\beqn
\tsigma^2 = - \left(\frac{\partial^2 \ln f}{\partial x^2} \right)^{-1}_{x=0} = - \frac{f(0)}{f''(0)}
\eeq
and the partition function gets approximated by the area under a Gaussian, which can be readily evaluated
as
\beq
\Xi(\sigma,\alpha) \approx \sqrt{2\pi} f(0) \tsigma = \frac 1{\sqrt{\pi\{1-\lambda z(\sigma)\}
\{1+\lambda b(\sigma)\}}},
\eeq
where
\bea
z(\sigma) &=& \sum_{l=1}^\infty z_l(\sigma)(l+1)\cr\cr
b(\sigma) &=& \sum_{l=1}^\infty z_l(\sigma)\left[\frac 23l^3 + 2l^2 + \frac 13 l-1\right].
\eea
This gives the average number of punctures as 
\beq
N = \frac{\partial \ln \Xi}{\partial\alpha} = \frac{\lambda\{z(\sigma)- b(\sigma)[1-2\lambda z(\sigma)]\}}
{2[1+\lambda b(\sigma)][1-\lambda z(\sigma)]}
\eeq
and, once again, the large $N$ limit requires $\lambda z(\sigma) \rightarrow 1^-$. 

If we set $1-\lambda z(\sigma) = \delta^2$ so that $\delta^2 \approx 0$, then to leading order it will 
be seen that
\beq
\delta \approx \frac 1{\sqrt{2N}} + \mathcal{O}(N^{-1}),~~ \lambda = e^\alpha \approx \frac 1z + 
\mathcal{O}(N^{-1})
\label{delta}
\eeq
and
\bea
\ln \Xi &\approx& \frac 12 \ln 2N,\cr\cr
a=\frac A{4\pi\gamma l_p^2 N} &\approx& - \frac{\partial\ln z(\sigma)}{\partial\sigma},
\label{alnz}
\eea
the last being identical to \eqref{qdef}. It can be used to determine $\sigma=\sigma(a)$ and $z=z(a)$ as 
before. Putting it all together, one finds that the entropy is now given to leading order in $N$ by
\beq
S(A,N) = \frac 12 \ln N + N[a\sigma(a)+ \ln z(a)].
\label{san2}
\eeq
The only difference between the entropy with the projection constraint and the entropy without the projection 
constraint in \eqref{san} is the coefficient of the logarithmic term, showing that the projection constraint 
has the only effect of subtracting $\frac 12\ln N$ from the entropy. This result was obtained in the canonical 
ensemble as well \cite{ours} where, there being no fluctuations permitted in $N$, a net negative logarithmic 
correction was found. The positive correction in the grand canonical ensemble results from fluctuations in the 
number of punctures and makes the entropy a concave function of the area. 

As before, the entropy is a complicated function of $A$ and $N$ but may be expressed, at fixed $\sigma=\sigma_0$, 
as a function of the area alone,
\beq
S(A) = \frac 1{\pi\gamma}\left[\sigma_0 + \frac{\ln z(a_0)}{a_0}\right]\frac A{4l_p^2} + \frac 12
\ln \frac A{4l_p^2} + \text{const.}
\eeq
For a vanishing chemical potential, the condition $z(\sigma_0)\approx 1$ still holds, according to 
\eqref{delta}. Thus one also obtains the same value of the Barbero-Immirzi parameter as in the absence
of the projection constraint.

As mentioned in the introduction, an interesting approach for the GCE was developed in \cite{bv11a} for the 
case of zero chemical potential. The partition functions worked with by the authors differ from ours in 
\eqref{partfunc}; in particular, for the DL counting scheme the authors evaluate
\beq
\Xi^\text{DL}(\sigma) = \frac 1{2\pi} \int_{-\pi}^\pi \frac {dk}{1-2\lambda(\alpha)\sum_{l=1}^\infty 
e^{-\sigma\sqrt{l(l+2)}} \cos k l} ,
\eeq
with $\lambda(\alpha)=1$ and for the ENP scheme,
\beq
\Xi^\text{ENP}(\sigma) = \frac 1{\pi} \int_{-\pi}^\pi \frac {\sin^2 k~ dk}{1-\lambda(\alpha)\sum_{l=1}^\infty 
e^{-\sigma \sqrt{l(l+2)}} \frac{\sin k(l+1)}{\sin k}}
\eeq
again with $\lambda(\alpha)=1$. The integrand of the first (DL) is also an unimodal, symmetric distribution 
centered at $k=0$ and can be treated in a manner identical to ours with the same result \eqref{san2} in the 
large $N$ limit. The integrand of the second (ENP) is bimodal, symmetric about $k=0$ and skew about each mode.
The absence of a logarithmic correction to the BH law in this case can be qualitatively understood if it can 
be shown that each mode has the effect of subtracting $\frac 12\ln N$ from the entropy. However, our methods 
are not adapted to this distribution and \cite{bv11a} appears to be the most effective approach to this problem 
at present.

\section{Equispaced Spectrum}

Given that the distribution function is not in fact Gaussian (if we assume that it is Gaussian and 
compute its kurtosis we should find an excess kurtosis of 3, which would contradict our assumption),
one may wonder how accurate the saddle point approximation is. At present we do not know of any way to 
formally estimate the error, so we will instead verify the leading order discussed above by using the 
equispaced spectrum mentioned earlier. For this spectrum, $\Xi$ can be evaluated exactly. Performing 
the sum in \eqref{partfunc}, with $z_l(\sigma) = e^{-\sigma (l+1)}$, we find
\beq
\Xi(\sigma,\alpha) =  \frac 1{2\pi} \int_{-\pi}^{\pi} \frac{dk}{1-\lambda\left(\frac{2\cos k-e^{-\sigma}}
{e^{2\sigma}-2e^\sigma \cos k + 1} \right)},
\eeq
and the integration may be carried out quite easily; one finds
\beq
\Xi(\sigma,\alpha) = \frac{AD\sqrt{\frac{C+D}{C-D}}+B\left(C+D-C\sqrt{\frac{C+D}{C-D}}\right)}{D(C+D)},
\eeq
where
\beq
A= e^{2\sigma}+1,~~ B = 2e^\sigma,~~ C = e^{2\sigma}+\lambda e^{-\sigma}+1,~~ D = 2(\lambda + e^\sigma),
\label{ABCD}
\eeq
provided that $C/D \in (-\infty,-1)\cup (1,\infty)$ for a real $\Xi(\sigma,\alpha)$. Since $C$ and $D$ 
are both positive we require $C/D >1$, or
\beq
\lambda z^{(0)}(\sigma) < 1.
\label{condition}
\eeq
Finally, using \eqref{ABCD} and after a little algebra, we obtain the exact partition function for 
the equispaced spectrum,
\beq
\Xi(\sigma,\alpha) = \frac{e^\sigma\left(\lambda e^\sigma
+ \sqrt{-4(\lambda+e^\sigma)^2+(1+\lambda e^{-\sigma}+e^{2\sigma})^2}\right)}{(e^\alpha+e^\sigma)
\sqrt{-4(\lambda+e^\sigma)^2+(1+\lambda e^{-\sigma}+e^{2\sigma})^2}}
\eeq
and \eqref{condition} will be recognized as the condition that the term under the radical is positive. 

Therefore, using the expression for $z^{(0)}$ in \eqref{zapprox}, we set, for a positive quantity 
$\delta^2$, which we do not as yet assume to be small,
\beq
\frac{\lambda(2-e^{-\sigma})}{(e^\sigma -1)^2 } = 1-\delta^2 \Rightarrow \lambda = e^\alpha = 
\frac{(e^\sigma -1)^2 (1-\delta^2)}{(2-e^{-\sigma})}.
\label{alphadelta}
\eeq
Now the chemical potential may be eliminated in favor or $\delta$ and the partition function written 
in terms of $(\sigma,\delta)$. We find
\beq
\Xi(\sigma,\delta) = \frac{(1+\delta^2)(2e^\sigma-1)\delta + e^{2\sigma}(e^\sigma-1)\sqrt{\frac{2e^\sigma-1}
{\delta^2(2e^{3\sigma}+3e^{2\sigma} -1)+4e^{3\sigma}}}}{\delta[e^{2\sigma}+(2e^\sigma-1)\delta^2]}
\eeq
and taking the derivative with respect to $\alpha$ at constant $\sigma$, 
{\it i.e.,}
\beq
\frac{\partial \ln \Xi(\sigma,\alpha)}{\partial \alpha}  = N = -\frac {1-\delta^2}{2\delta} \frac
{\partial \ln\Xi(\sigma,\delta)}{\partial\delta},
\label{derd}
\eeq
recovers $N$.

In the thermodynamic limit we obtain that either 
$e^\sigma\approx1$ or $\delta\approx0$. For $e^\sigma\approx 1$, we will see that $a\rightarrow \infty$, 
which we discard as being unphysical since it implies that $N\rightarrow 0$ for a finite $A$. Therefore 
we take $\delta \approx 0$ and consider a small $\delta$ expansion of $\ln\Xi$:
\beq
\ln \Xi = \ln \left(\frac{(e^\sigma-1)\sqrt{e^{-3\sigma}(2e^\sigma-1)}}{2\delta}\right) + 
{\mathcal{O}}(\delta)
\eeq
so
\beq
N \approx -\frac 1{2\delta} \frac{\partial\ln\Xi(\sigma,\delta)}{\partial \delta} \approx \frac 
1{2\delta^2} + \frac{\sqrt{z^{(0)}}}{\delta} + {\mathcal{O}}(\delta^0),
\eeq
which gives
\beq
\delta \approx \frac 1{\sqrt{2N}} + {\mathcal{O}}(N^{-1})
\eeq
and, according to \eqref{alphadelta}, 
\beq
\lambda \approx \frac{(e^\sigma -1)^2 }{2-e^{-\sigma}} + \mathcal{O}(N^{-1}) = \frac 1{z^{(0)}} + 
\mathcal{O}(N^{-1}),
\label{fug}
\eeq
both of which results agree with \eqref{delta}. Again, if we consider the average area, we find
\beq
\frac A{4\pi \gamma l_p^2} = - \frac{\partial \ln \Xi(\sigma,\alpha)}{\partial\sigma} \approx 
\frac{1-3e^\sigma + 4e^{2\sigma}}{2(e^\sigma-1)(2e^\sigma-1)\delta^2} - \frac{e^{-\sigma/2}
(1-3e^\sigma + 4e^{2\sigma})}{(e^\sigma-1)^2\sqrt{2e^\sigma-1}~\delta} + {\mathcal{O}}(\delta^0)
\eeq
which, to leading order becomes
\beq
a~ \stackrel{\text{def}}{=}~ \frac A{4\pi\gamma l_p^2N} \approx \frac{1-3e^\sigma + 4e^{2\sigma}}
{(e^\sigma-1)(2e^\sigma-1)} = - \frac{\partial \ln z^{(0)}}{\partial \sigma},
\label{asig}
\eeq
as given in \eqref{alnz}. Solving for $\sigma$ gives
\beq
e^\sigma = \frac{3(1-a)\pm\sqrt{(a-1)(a+7)}}{4(2-a)}.
\eeq
Now both solutions are real only so long as $a\geq 1$. The solution with the positive sign is 
monotonically increasing from zero to $1/2$ with $a \in [1,\infty)$. This solution is unacceptable 
since $e^\sigma$ must be $\geq 1$. Thus we conclude that
\beq
e^\sigma = \frac{3(1-a)-\sqrt{(a-1)(a+7)}}{4(2-a)}
\eeq
and $a>2$. As $e^\sigma \rightarrow \infty$ we see that $a \rightarrow 2^+$ and $\lambda \rightarrow \infty$. 
On the other hand, as $e^\sigma \rightarrow 1^+$ then we have $a \rightarrow \infty$ or $N\rightarrow 0$ and 
$\lambda \rightarrow 0$. Since both \eqref{delta} and \eqref{alnz} are satisfied, the entropy will be given 
by \eqref{san2}, with $z(\sigma)$ replaced by $z^{(0)}(\sigma)$.

\section{Conclusion and outlook}

In this paper we have followed up on work begun in \cite{ours} by providing a detailed analysis of the 
entropy of loop black holes in the grand canonical ensemble (GCE), employing the counting scheme of GM.
As pointed out in the introduction and in \cite{Ghosh}, the grand canonical ensemble is probably the one 
that is most relevant to the description of physical processes, such as evaporation, which involve changes 
in the horizon area. 

We observe that in the thermodynamic limit the ratio of the average area to the average number of 
punctures emerges as a fundamental variable that controls both the ``temperature'' (in the area 
ensemble) and the chemical potential of the black hole. The entropy is in general a complicated 
function of the average number of punctures and the average area, and the BH law is recovered only 
for isothermal processes (equivalently, processes with constant chemical potential).

We have also shown that a positive logarithmic correction to the BH entropy is obtained in the GM counting 
scheme and it is independent of any particular choice of the chemical potential. The same can be expected 
for the DL counting scheme, as we have argued in Section III, since it is a unimodal, symmetric distribution 
centered at $k=0$. The source of the positive correction 
was identified as the fluctuations in the number of punctures and the projection constraint was shown to 
decrease the number of configurations by subtracting $\frac1{2}\log{A}$ from the entropy for an isolated 
horizon. The resulting concavity of the entropy as a function of the area signals the stability of the black 
hole in the area ensemble, \cite{Barbero},\cite{Gaur}, however such a stability cannot be deemed a ``thermal'' 
stability so long as a rigorous relation motivated from LQG linking the area of an isolated horizon and 
some quasi-local energy function does not exist. 

An attempt to link the two was made by Ghosh and Perez \cite{Ghosh} who showed that the energy 
associated with an isolated horizon by a preferred family local observers at a proper distance 
$l$ from it is proportional to its area. The constant of proportionality was determined to be the 
local surface gravity measured by the locally non-rotating, stationary observer and Ghosh and Perez 
were able to work in the microcanonical and canonical energy ensembles. Pranzetti \cite{pran12} 
extended their construction to the grand canonical ensemble and by matching the description of weakly 
dynamical horizons \cite{weakdyn1, weakdyn2} with the local statistical description was able 
to obtain a temperature regulating the exchange of energy between the bulk and the horizon in terms 
of the local surface gravity of the horizon. This procedure also involved singling out a physical 
time parameter, equivalently a preferred family of observers, with respect to which the evolution 
of the boundary states was described. 

These works are important steps in the direction of a fully quantum derivation of the Hawking radiation 
from black holes. We feel that rigorously connecting different statistical-thermodynamic processes 
with families of observers via our analysis will give thermo-statistical meaning to the choice of vacuum 
in the semi-classical theory and is the natural next step in this analysis. We will report on the results of 
our findings elsewhere.

\bigskip\bigskip

\noindent{\bf ACKNOWLEDGEMENTS}
\bigskip

\noindent KL wishes to thank S. Sahu for his useful comments. This research was supported in part by 
the Templeton Foundation under Project ID $\#$ 20768.

\end{document}